\documentclass[taps,prl,superscriptaddress,preprint,amsmath,amssymb,english]{revtex4}
\usepackage[T1]{fontenc}
\usepackage[latin9]{inputenc}
\usepackage{hyperref} 
\usepackage{graphicx}

\makeatletter
\def\ket#1{\left| #1\right>}

\makeatother
\usepackage{babel}

\begin{document}

\title{Non-equilibrium Fractional Quantum Hall state of light }

\author{Mohammad Hafezi}
\email{hafezi@umd.edu}
\affiliation{Joint Quantum Institute, NIST/UMD, College Park MD 20742}
\author{Mikhail D. Lukin}
\affiliation{Physics Department, Harvard University, Cambridge, MA 02138}
\author{Jacob M. Taylor}
\affiliation{Joint Quantum Institute, NIST/UMD, College Park MD 20742}

\begin{abstract}
We investigate the quantum dynamics of systems involving small numbers of strongly interacting
photons. Specifically, we develop an efficient method to investigate
such systems when they are externally driven with a coherent field. Furthermore, we show how to quantify the many-body quantum state of light via correlation functions. Finally,  we apply this method to two strongly interacting cases: the
Bose-Hubbard and fractional quantum Hall models, and discuss an implementation of these ideas in atom-photon system.\end{abstract}
\maketitle

Strongly interacting photonic systems provide new avenues for examining quantum simulation, topological quantum computation, and many-body states of matter
~\cite{Angelakis:2007,Greentree:2006,Hartmann:2006,Chang:2008,Schmidt:2010}. Inspired by analogies to electronic systems, these studies focus on ground state properties. However, a photonic system
is naturally an open driven system. At the same time, this is in stark contrast to ultracold atomic systems, which have been extensively studied in the context of many-body physics; in most cases atoms are trapped in a potential and the particle number is conserved within the trapping time \cite{cooper,dalibard,bloch}.
Therefore, the most relevant approach to understand and manipulate many-photon states
involves understanding the non-equilibrium dynamics
in such systems \cite{Carusotto:2009,Tomadin:2010ul,Hafezi:2011p36115,Nunnenkamp:2011dm,Nissen:2012,schiro:2012}. For example, in a one-dimensional system strong interaction between photons leads to their fermionization, which can be probed in the output correlation functions of an externally driven system, both in a discrete array \cite{Carusotto:2009} and in the continuum limit \cite{Hafezi:2011p36115}. Another unique property of photonic system is the lack of chemical potential, in contrast to other bosonic systems. In particular, these differences raises key questions: given the presence of photon loss, how does one prepare a photonic state with many-body
features? What is the manifestation of important properties such as incompressibility and collective effects, when the system is coherently driven with a laser field rather than coupled to a thermal bath?

In this article, we address these questions by studying a driven system of strongly
interacting photons and evaluating physical observables that
display quantum many-body signatures. We focus on
a two-dimensional lattice of interacting photons with an effective gauge
field \cite{Koch:2010, Hafezi:2011delay,Hafezi:2011ui,Umucalilar:2011}. In the presence of strong interaction (nonlinearity) on each site, the system maps into the bosonic fractional quantum Hall  \cite{Cho:2008,Umucalilar:2011b,Hayward:2012} model. Such nonlinearities have been experimentally shown at the single site (resonator), both for
optical \cite{Kimble:2005,Englund:2007,Srinivasan:2007p4061} and
microwave \cite{Schoelkopf:2008p8712,You:2011jj} photons. We demonstrate
that by weakly driving the system,  a few photon Laughlin
state can be prepared. We introduce experimentally-relevant observables such as the correlation function
of the zero-mode (the common-mode) to investigate the response of the system. Furthermore, we present a scheme to adibatically prepare such state using many-photon Fock state and compare it to a driven scheme.

The key idea underlying our approach to the driven scheme is to generalize the theoretical technique of  weakly driven cavity-QED system (involving one atom interacting with one photon) to the many sites and many-photon regime. We follow Carmichael et al. \cite{Carmichael:1991} who showed that when an optical resonator with strong nonlinearity is weakly driven, one can truncate the Hilbert space up to two-excitation states, reduce the exact master equation description to an effective Schr\"{o}dinger equation description. Most importantly,  the ``quantum jumps'' do not contribute in  the correlation functions.  In particular, while the one-photon state is intact under nonlinearity, the two-photon component exhibits bunching or anti-bunching features. Similarly, in a system of many sites, the photonic state at each photon-number manifold reorganize themselves according to the interaction. Ignoring the quantum jumps has a significant benefit which allows the investigation of larger systems and avoids finite size effects in numerical simulations. In a dilute lattice with $N_{\phi}$ magnetic flux quanta and strong interaction for a fixed number of bosonic particles $(N_{ph})$, the system 
is expected to have fractional quantum Hall states (Laughlin-type) at filling factors $\nu=N_{ph}/N_{\phi}=1/2,1/4,...$
 \cite{sorensen,Hafezi:PRA2007}. We demonstrate that when an optical system is driven with a weak coherent field, which has Poissonian distribution of photon number, the system forms Laughlin state in a photon-number  ($N_{ph}$) manifold which corresponds to the bosonic Laughlin filling factors, at specific pump frequencies. We show that measuring the $N_{ph}$-body correlation function reveals the existence of such state. Furthermore, we present an alternative adiabatic method to prepare such a state for larger photon number and compare the two methods. While our results are general and can be implemented in various photonic systems, we focus one a physical implementation of these ideas with coupled optical resonators.

 \section{ Driven photonic quantum Hall model on a lattice}
 
We consider a 2D interacting photonic system which has the Hamiltonian ($\hbar = 1$):
\begin{eqnarray}
H_{\rm sys} & = & -J\sum_{x,y}\hat{a}_{x+1,y}^{\dagger}\hat{a}_{x,y}e^{i2\pi\alpha y}+\hat{a}_{x,y}^{\dagger}\hat{a}_{x+1,y}e^{-i2\pi\alpha y}\nonumber \\
 &  & \ +\  \hat{a}_{x,y+1}^{\dagger}\hat{a}_{x,y}+\hat{a}_{x,y}^{\dagger}\hat{a}_{x,y+1}+H_{\rm free} + H_{\rm int}.\label{eq:hamiltonian_system}
\end{eqnarray}
where $a_{x,y}^{\dagger}$ is the creation operator at site $(x,y)$, $J$ is tunneling rate between resonators, $\alpha$ the effective
magnetic flux per plaquette (total magnetic flux is $N_{\phi}=\alpha N_{x}N_{y}$), and $H_{\rm free} = \sum_{x,y} \omega_0 a_{x,y}^\dag a_{x,y}$.
We take an on-site interaction term of the Kerr-type:
$H_{\rm int}=U\hat{a}_{i}^{\dagger}\hat{a}_{i}(\hat{a}_{i}^{\dagger}\hat{a}_{i}-1)$
where the index stands for the site $i=(x,y)$. In the absence of
the magnetic field $(\alpha=0)$, the Hamiltonian describes the Bose-Hubbard
model, and can be implemented in an array of coupled optical resonator ~\cite{Angelakis:2007,Greentree:2006,Hartmann:2006}.  The non-zero magnetic field can be synthesized using an imbalance in the optical paths that connect resonators. We return to the discussion of implementation of such Hamiltonian extending the scheme proposed in Ref. \cite{Hafezi:2011delay} later in the article. 

To include loss and driving, we use the stochastic wave-function approach \cite{Carmichael:1991,Dalibard:1992,carmichael:book}.  The coherent drive is applied uniformly; its effects and that of the associated loss can be described by the non-Hermitian term
$H_{\rm pump}=\sum_{i}\kappa\beta(e^{i\omega_{p}t}\hat{a}_{i}+e^{-i\omega_{p}t}\hat{a}_{i}^{\dagger})-i\kappa\hat{a}_{i}^{\dagger}\hat{a}_{i}$,
where $\kappa$ is the coupling rate to the resonators, $\beta$ is the amplitude and $\omega_p$ the frequency of the drive field.  In the rotating frame of the pump field, the
effective Hamiltonian of the driven system is:
\begin{equation}
H_{\rm eff} =  H_{\rm sys}+\kappa\beta\sum_i(\hat{a}_{i}+\hat{a}_{i}^{\dagger}) - (\Delta+i\kappa)\sum_{i}\hat{n}_{i}\label{eq:eff_ham}
\end{equation}
where the pump detuning $\Delta = \omega_p - \omega_0$ takes the form of a chemical potential. Since the system is open, in the absence of the pump ($\beta=0$), the system will be in the vacuum state.

We generalize the quantum-jump picture for evaluating the correlation functions \cite{Carmichael:1991} to many-photons and many-modes. The evolution of the system is governed by the effective Hamiltonian (Eq.\ref{eq:eff_ham}) and the corresponding quantum jump operators ($\hat{a}_i$). In particular, in the weakly excited system ($\beta\ll1$), the metastable state of the system can be perturbatively written as: 
\begin{equation}
|\Psi\rangle\simeq| 0\rangle+\mathcal{O}(\beta)|1\rangle+\mathcal{O}(\beta^{2})|2\rangle+...\mathcal{O}(\beta^{n})|n\rangle+...
\label{eq:psi}
\end{equation}
where $|n\rangle=\sum c_{i_1...i_n}\hat{a}^\dagger_{i_1}...\hat{a}^\dagger_{i_n}|0\rangle$ represents a state in the $n$-photon manifold of
the lattice system. This state is the eigenstate of $H_{eff}$ with the smallest imaginary eigenvalues, i.e. it is mostly the vacuum state.  All other states  have at least one photon, and therefore, they decay rapidly into this state. When a photon decays from any site, the system undergoes a quantum jump. These jumps occur at a rate $~\kappa \mathcal{O}(\beta^2)$, and the system takes a state of the form: $|\Psi_i'\rangle=\hat{a}_i|\Psi\rangle/(|\hat{a}_i|\Psi\rangle|)\simeq | 0\rangle+\mathcal{O}(\beta)|1'\rangle+\mathcal{O}(\beta^{2})|2'\rangle+...\mathcal{O}(\beta^{n})|n'\rangle+...$. Similarly, the system can undergo a two-photon jump with a slower rate $\kappa \mathcal{O}(\beta^{4})$. Since the system is continuously pumped, it is restored back into the steady state with a relatively fast rate ($\kappa$). Therefore, the density matrix of the system can be  formally  written as: $\rho=|\Psi\rangle \langle\Psi|+ \mathcal{O}(\beta^{2}) \rho_1 +\mathcal{O}(\beta^{4}) \rho_2+...$, where $\rho_j$ stands for the density matrix after ``j'' consecutive jumps. For a single jump we have, $\rho_1=(1/\sum_i \langle \Psi | \hat{a}^\dagger_i\hat{a}_i|\Psi\rangle)\sum_i \langle \Psi | \hat{a}^\dagger_i\hat{a}_i|\Psi\rangle |\Psi_i'\rangle \langle \Psi_i'|$.
Now, we evaluate the n-body correlation function of an arbitrary operator $\hat{d}$ which is a linear superposition of the site operators ($\hat{a}_i$). In particular, we are interested in $G^{(n)}=\langle\rho \hat{d}^{\dagger n} \hat{d}^n \rangle$. Using the above picture, this correlation function can be perturbatively written in powers of pump amplitude:  
\begin{equation}
G^{(n)}=\mathcal{O}(\beta^{2n}) \langle n| \hat{d}^{\dagger^n} \hat{d}^n |n \rangle+\mathcal{O}(\beta^{2n+2}) \langle n'| \hat{d}^{\dagger n} \hat{d}^n |n' \rangle+...
\end{equation}
Therefore, if we are interested in the $n$-photon manifold, the metastable state $|\Psi
\rangle$ is sufficient for evaluation of any n-body correlation function and the corrections due to quantum jumps can be ignored. 

In particular, for a two-particle case, we define the two-body observables to characterize the deviation from
the classical regime. For a single resonator this deviation is characterized
by the equal time second-order correlation function as $g^{(2)}=\langle \hat{a}_{i}^{\dagger2}\hat{a}_{i}^{2} \rangle/\langle \hat{a}_{i}^{\dagger}\hat{a}_{i}\rangle^{2}$.
This quantity is useful in characterization of cavity QED experiments. However, this observable can
not encapsulate the collective effects in the system. In particular,
in the presence of strong interaction $(U\gg J)$, such a quantity is
always less than one regardless of the collective features of the
entire system. Instead, we consider a collective observable which is the second-order correlation function of the common-mode
$(\hat{b}^{\dagger}=\frac{1}{\sqrt{N}}\sum_{i}\hat{a}_{i}^{\dagger})$: 
$g_{CM}^{(2)}=\frac{\langle\hat{b}^{\dagger2}\hat{b}^{2}\rangle}{\langle\hat{b}^{\dagger}\hat{b}\rangle^{2}}.$ This observable is particularly interesting since we are exciting
all the resonators the same way, and therefore, this mode is primarily
excited. Such quantity can be obtained by measuring the 2nd-order
correlation function of the far-field light emitted from all the resonators. In the context of ultra cold gases confined in optical cavities,  optical correlation functions can reveal many-body physics of the atomic system \cite{mekhov:phys, chen, mekhov}.

To numerically find $\ket{\Psi}$, we consider a truncated
Hilbert space corresponding to at most a few particles and find the eigenstate
of $H_{\rm eff}$ with the smallest imaginary part of its eigenvalue.  
Note that in contrast to grand canonical
ensemble -- where we minimize ($\hat{H}-\mu\hat{N}$) -- here we find
the steady state of the system as a function of the pump field. Such
approach allows us to consider larger lattices which are otherwise inaccessible with the density matrix
approaches \cite{Umucalilar:2011b}.  

\section{Overlap with Laughlin wavefunction and correlation functions}

Using the technique described above, we study the driven system of
interacting photons with the Hamiltonian of Eq.(\ref{eq:hamiltonian_system}).
First, we consider the case of hard-core bosons ($U\gg J$), and investigate
the response of the system as a function of the pump field frequency ($\Delta$). For simplicity, we only consider the case
of $\nu=1/2$. The input field consists of a Poisson distribution
of photons. When photons are injected at the frequency corresponding
to the Laughlin state at the $N_{ph}$-photon manifold, photons reconfigure
themselves and form a wave function which corresponds to the Laughlin
state. The remarkable overlap of this photonic state with the Laughlin
wave function in the $N_{ph}$-photon manifold is shown in Fig.1(a).
Note the frequency required to be resonant with the Laughlin state
is at the vicinity of the free photon state (Hofstadter's spectrum).
In the limit of large system $(N_{x}N_{y}\rightarrow\infty)$, and
dilute magnetic field $(\alpha\ll1)$, these two frequencies coincide
since the Laughlin state is the exact ground state of the Hamiltonian
in the continuum limit. For numerical simulations, we have used the
discrete version of the Laughlin wave function on the lattice with
torus boundary condition \cite{Hafezi:PRA2007}.

Around the resonance, we observe the suppression
of the correlation function of the common-mode. The reason behind this suppression is that the external pump is coupled differently to the single particle manifold and $N_{ph}$-photon manifold, corresponding to the Laughlin filing factor. We note that the energy of the single particle state and the Laughlin state per particle is exactly equal to each other in the continuum limit, and the previously reported discrepancy is due to the finite size effect \cite{Umucalilar:2011}.  The direct experimental verification of the
Laughlin overlap is a difficult task which requires number post-selection
$(N_{ph})$ and state tomography in a Hilbert space with dimension $\left(\begin{array}{c}
N_{x}N_{y}\\
N_{ph}
\end{array}\right)$. However, the common-mode correlation function
can be obtained by using conventional quantum optics measurements.  

Now, we relax the hard-core constraint and investigate the same observables.
In the weak interaction limit, the system approaches the classical
response, as shown in Fig. 1(b). In the absence of interaction, using
transport measurements -- varying the pump frequency and measuring
reflection/transmission-- one recovers the Hofstadter's butterfly
spectrum \cite{Hafezi:2011delay}, but regardless of the pump frequency, the correlation function remains equal to one. Similar behavior was observed for $N_{ph}=3$, as shown in Fig.1 (c,d).

\begin{figure}[t]
\includegraphics[width=0.70\textwidth]{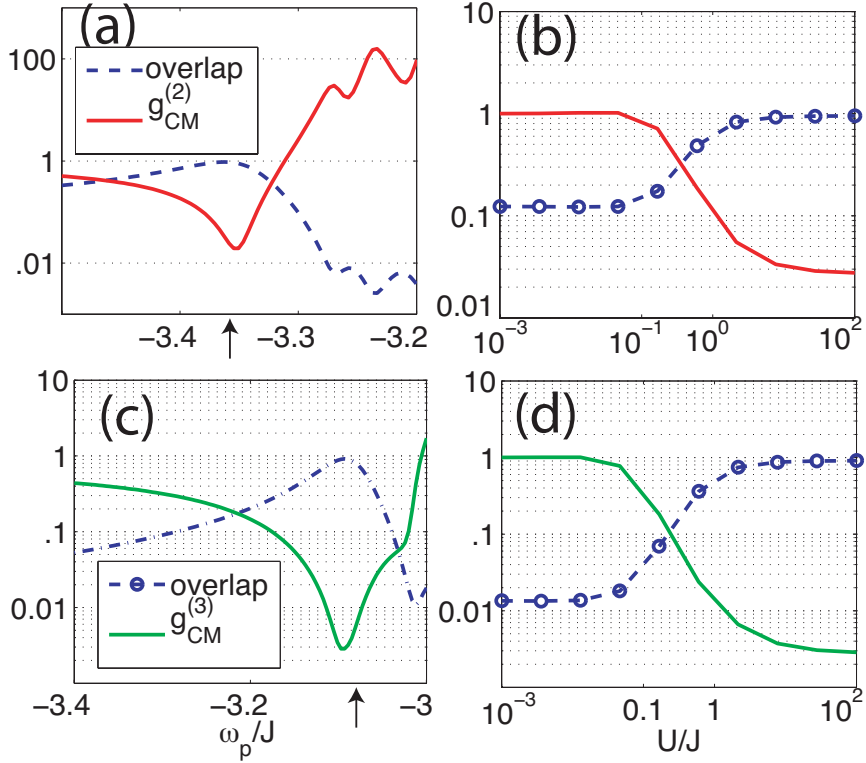}\caption{\label{fig:FQH_fr_U} Overlap with the Laughlin wave function ($\nu=1/2$), and the correlation function of
the zero mode ($g_{CM}^{(2)}$) are shown as a function of: (a) the
pump frequency for hard-core bosons (b) the interaction strength for
$\Delta=-3.36J$, as shown by an arrow on (a).  We have evaluated the overlap with the Laughlin
function in $N_{ph}=2$ manifold. The total magnetic flux is $N_{\phi}=4$. (c,d) are similar to (a,b) for $N_{ph}=3$, $N_{\phi}=6$, $\Delta=-3.095J$ and the corresponding correlation function ($g_{CM}^{(3)}$). All the simulations are performed for a 6x6 lattice, torus boundary condition, and
the maximum number of photon is 3. $\kappa=.01J,\beta=0.01$. All calculated quantities are dimensionless.}
\end{figure}

In other to clarify the connection between  zero mode correlation and the collective nature of the system response, we investigate the driven photonic Bose-Hubbard model (Fig.\ 2) \cite{Tomadin:2010ul,Nissen:2012}. In the limit of weak interaction, the system behaves
classically and the correlation function approaches that of a coherent state, i.e. $g_{CM}^{(2)}=1$,
as shown in Fig. 2(b). However, in the strong interaction limit ($U\gg J$) the system exhibits significant deviation from a classical state \cite{Nissen:2012}. In contrast, to the previous works  \cite{Tomadin:2010ul,Nissen:2012}, we focus in the weakly driven regime, and therefore, we expect that the system to be in the superfluid state and the correlation function to be equal to one. This deviation is due to the finite size of the system and can be
understood in the following way: the system is weakly driven and manifolds
with large number of photons are weakly populated.
Therefore, the effective filling factor $\langle n_{tot}\rangle/(N_{x}N_{y})$
is small and in the presence of a non-zero interaction, one expects
the system to be in a superfluid regime.
However, due to finite size of the system, the common-mode is not completely harmonic and the two-photon resonance is slightly shifted. This leads to a deviation of the correlation function from unity; using the single-mode approximation, we get an estimate $g_{max}^{(2)}=1+\left(\frac{\delta U}{\kappa}\right)^{2}$, where $\delta U$ is the nonlinear shift, i. e., the difference between half of the  two-photon state  energy and the single-photon state energy. Such nonlinearity decrease with the system size, in direct analogy to spin-boson transformation (Holstein\textendash{}Primakoff) of the Dicke-model, where the residual nonlinearity disappears in the limit of large spins.

\begin{figure}[t]
\includegraphics[width=0.50\textwidth]{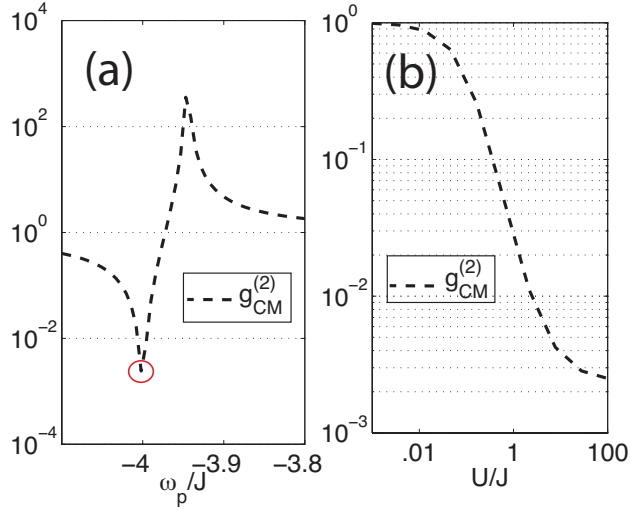}\caption{\label{fig:BH_fr_U} The correlation
function of the zero mode ($g_{CM}^{(2)}$) are shown as a function
of: (a) the pump frequency for hard-core bosons (b) the interaction
strength for $\Delta=-4.0J$ (where $g_{CM}^{(2)}$ is minimum, as shown with a red circle on (a)).
All simulations are performed  for a 6x6 lattice, torus boundary condition and
the maximum number of photon is 3. $\kappa=0.002J,\beta=0.01$. }
\end{figure}

We numerically verify such statement by evaluating the correlation function ($g_{CM}^{(2)}$) as a function of the system size. In the Bose-Hubbard model, as the system size increases, the correlation function  $g_{CM}^{(2)}$
approach the classical limit, i.e. unity, as shown in Fig. 3(a), while the correlation function of individual sites is equal to zero. The green curve shows the numerical estimate based on the nonlinearity
between one- and two-photon manifold lowest energies, which diminishes
as the system size increases. In contrast, in the FQH model, $g_{CM}^{(2)}$ remains constant as the system
size changes, as shown in Fig. 3(b). Note that the overlap with the
Laughlin wave function is also constant and remains close to unity. We have also performed numerical simulation for two-point correlation function $g(i,j)=\langle \hat{a}_{i}^{\dagger}\hat{a}_{j}^{\dagger}\hat{a}_{j} \hat{a}_{i} \rangle$ projected into the $N_{ph}$-photon manifold, and the results agrees with two-point correlation of the Laughlin state. Note that in the general case of $n$-photon FQH state, one should measure $n$-body correlation function  $G^{(n)}=\langle\rho \hat{d}^{\dagger n} \hat{d}^n \rangle$, as introduced earlier. Such correlation function can be measured using a modified Hanbury Brown-Twiss setup \cite{kartner93}:  the photonic mode $\hat{d}$ is collected, the light passes through $n$ beam splitters and then the state is detected using $n$ photodetectors. 

\begin{figure}[t]
\includegraphics[width=0.60\textwidth]{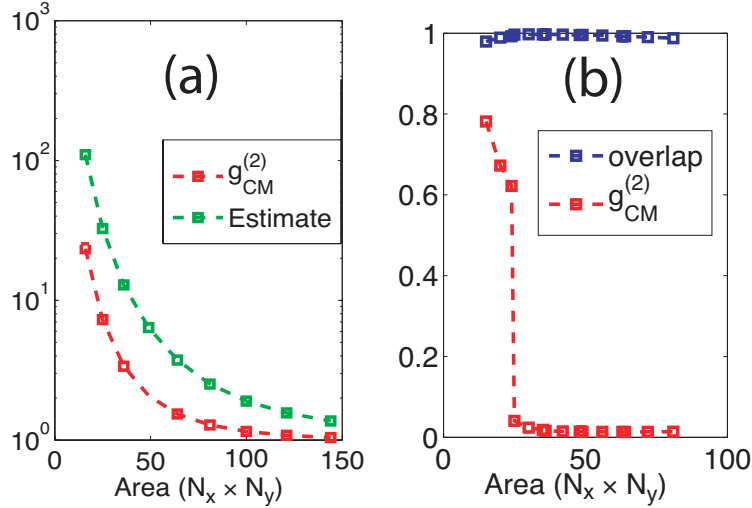}\caption{\label{fig:scaling} The correlation
function of the zero mode ($g_{CM}^{(2)}$) are shown as a function
of the system area ($N_{x}\times N_{y}$) for: (a) Bose-Hubbard and (b) Fractional Quantum Hall models. The overlap with the Laughlin wavefunction is shown in (a). The correlation function estimate is based on single-mode approximation (see  text). $N_{\phi}=2,\kappa=0.04J,\beta=0.01$ and the system is truncated at three photons.}
\end{figure}

\section{Possible implementations and outlook}

Now we discuss the implementation of the Hamiltonian in Eq.(\ref{eq:hamiltonian_system})
and the conditions to observe fractional quantum Hall states of photons.
Recently, there have been several proposals to implement the
artificial magnetic fields for photons \cite{Hafezi:2011delay,Koch:2010,Umucalilar:2011,Cho:2008,Hafezi:2011ui}
and various means to achieve strong interaction in coupled resonators
systems \cite{Greentree:2006,Hartmann:2006,Angelakis:2007}. Here,
we focus on the proposal in Ref.\cite{Hafezi:2011delay}  which does
not require time-reversal symmetry breaking for the implementation
of the magnetic field. Strong photon-photon interaction --which can
lead to photon blockade -- can be mediated by coupling emitters (e.g.,
atoms \cite{Bajcsy:2009p6498}, quantum-dots \cite{Fushman:2008},
Rydberg states \cite{Gaetan:2009,Urban:2009,Peyronel:2012} for optical photons and
Josephson junctions for microwave photons \cite{Lang:2011})
to the resonators. 

\begin{figure}
\includegraphics[width=0.70\textwidth]{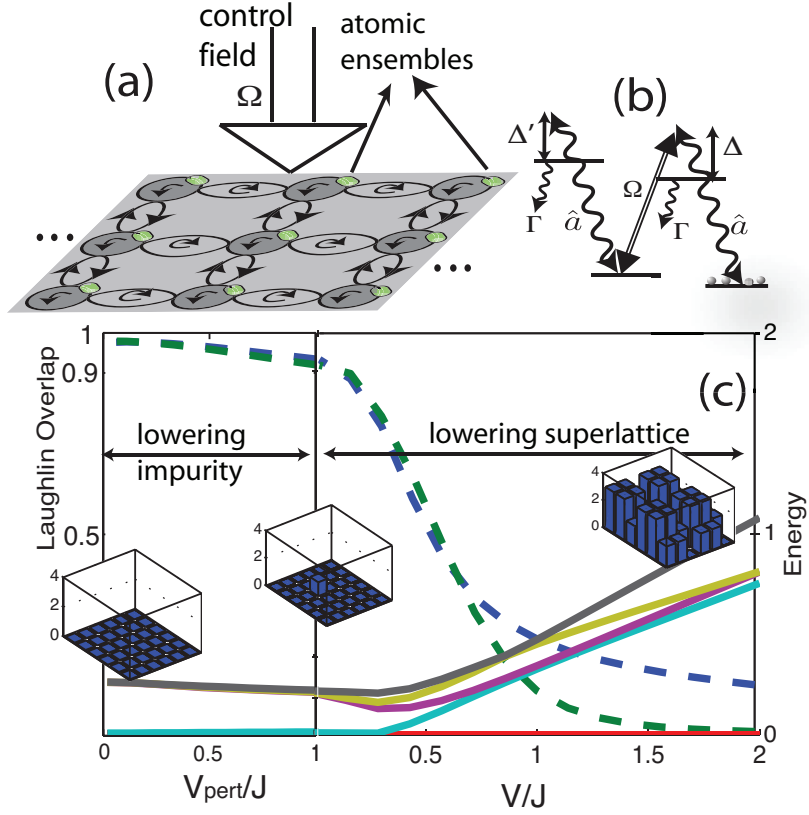} \caption{\textbf{Adiabatic preparation photonic Laughlin states:} (a) Atomic
ensembles are coupled to resonators to mediate interaction. A control
field couples internal levels of the atom, shown in (b), and provides
on-site interaction for photons \cite{Hartmann:2006}. (c) Overlap
of the two lowest states with Laughlin wave function (energy levels
relative to the ground state) are represented by dashed (solid) lines,
respectively. The procedure to make a Laughlin state: (i) Create $N_{p}$
photons in the whole system (e.g., by using lambda systems inside
the resonator), at this stage $\alpha$ is set to be zero. (ii) Make
a $N'_{x}\times N'_{y}$ superlattice potential V (e.g., by detuning
selected resonators) such that the ground state gets to the first
Mott insulator ($N_{ph}=N'_{x}N'_{y}$). (iii) Turn on a single-site potential
$V_{pert}$ by detuning a cavity (in this case (x,y)=(3,3)). (iv) Turn on the magnetic field to the desired value $\alpha=N_{ph}/(\nu N_{x}N_{y})$.
(v) Melt the Mott insulator by lowering the superlattice potential
strength to zero. (vi) Lower the single-site potential. Three snapshots
of lattice potential are shown at (iv) and the end of (v) and (vi)
steps, respectively. The impurity potential splits the ground state
degeneracy of the Laughlin state on the torus boundary condition \cite{Hafezi:PRA2007}
and prevents level crossing and sharp changes in the overlap. \label{fig:interaction} }
\end{figure}

Besides the driven method to reach fractional quantum Hall state that
we discussed above, one can also prepare a Laughlin state by adiabatically
melting a Mott-insulator of photons, similar to the atomic method
discussed in Ref. \cite{sorensen}, as described in Fig.~\ref{fig:interaction}.  However, this requires both preparation of $N_{ph}$ Fock states and photon lifetimes long enough to allow for the melting to be adiabatic, making the coherent drive approach preferable. Note that one might be able to use the nonlinearity of the system itself to prepare the $N_{ph}$ Fock states of photons \cite{Pohl:2010}.

Regardless of the preparation method, coupling atoms to the photonic system introduces loss which can be reduced by detuning the cavity resonance from the emitter transitions ($\Delta,\Delta'\gg\Gamma$).  As an example case, one can use an ensemble of N-level atoms to mediate onsite two-body interaction of the Kerr-type (Fig.\ref{fig:interaction}(b))\cite{Hartmann:2006}, which still preserves the propagation direction (clockwise or counterclockwise) used in Ref.~\cite{Hafezi:2011delay}.  In this approach, the optical cavity and ensemble enter into a slow-light regime, where the excitations are dark state polaritons~\cite{Fleischhauer:2000kx} $\hat{\Psi}_{x,y} \propto \Omega \hat a_{x,y} - g \sqrt{N} \hat S_{x,y}$, where $\Omega$ is the pump field, $g$ is the vacuum Rabi coupling, $N$ is the number of ensemble atoms, and $\hat S_{x,y}$ is the spin-wave operator describing coherence between two atomic states $\ket{a}$ and $\ket{c}$ (from Fig.~\ref{fig:interaction}(b)).  These bosonic excitations lead to an overall increase of dynamical timescales by $\eta = c/v_g \gg 1$, the ratio between the speed of light and group velocity for the dark state polariton, but they can also interact via a self-Kerr interaction with state $\ket{d}$ \cite{andre:NLO} .  For observing a Laughlin state and having a finite gap, the effective interaction between photons ($U\simeq g^{2}/\Delta'$) should be at least comparable to the tunneling rate J \cite{Hafezi:PRA2007}. These conditions can be satisfied for systems with a large Purcell factor ($g^{2}/\kappa\Gamma\gg1$). The same criterion applies to implementation of such scheme in the microwave domain.

In conclusion, we have shown that driven strongly interacting photons
exhibits interesting many-body behaviors and FQH state of photons and their incompressibility
can be probed by using conventional optical measurement techniques.
Investigation of other many-body signatures of these states such as
their topological properties and fractional statistics and preparation of photonic many-body state with reservoir engineering \cite{Diehl:2008} can be the subject of further research.

This research was supported by the U.S. Army Research Office MURI award W911NF0910406,  NSF through the Physics Frontier Center at the Joint Quantum Institute, CUA, Packard, Darpa and AFOSR MURI. We thank E. Demler and I. Carusotto for fruitful discussions and E. Goldschmidt and S. Polyakov for critical reading of the manuscript.


\begin{thebibliography}{39}
\expandafter\ifx\csname natexlab\endcsname\relax\def\natexlab#1{#1}\fi
\expandafter\ifx\csname bibnamefont\endcsname\relax
  \def\bibnamefont#1{#1}\fi
\expandafter\ifx\csname bibfnamefont\endcsname\relax
  \def\bibfnamefont#1{#1}\fi
\expandafter\ifx\csname citenamefont\endcsname\relax
  \def\citenamefont#1{#1}\fi
\expandafter\ifx\csname url\endcsname\relax
  \def\url#1{\texttt{#1}}\fi
\expandafter\ifx\csname urlprefix\endcsname\relax\def\urlprefix{URL }\fi
\providecommand{\bibinfo}[2]{#2}
\providecommand{\eprint}[2][]{\url{#2}}

\bibitem[{\citenamefont{Angelakis et~al.}(2007)\citenamefont{Angelakis, Santos,
  and Bose}}]{Angelakis:2007}
\bibinfo{author}{\bibfnamefont{D.}~\bibnamefont{Angelakis}},
  \bibinfo{author}{\bibfnamefont{M.}~\bibnamefont{Santos}}, \bibnamefont{and}
  \bibinfo{author}{\bibfnamefont{S.}~\bibnamefont{Bose}},
  \bibinfo{journal}{Physical Review A} \textbf{\bibinfo{volume}{76}},
  \bibinfo{pages}{31805} (\bibinfo{year}{2007}).

\bibitem[{\citenamefont{Greentree et~al.}(2006)\citenamefont{Greentree, Tahan,
  Cole, and Hollenberg}}]{Greentree:2006}
\bibinfo{author}{\bibfnamefont{A.~D.} \bibnamefont{Greentree}},
  \bibinfo{author}{\bibfnamefont{C.}~\bibnamefont{Tahan}},
  \bibinfo{author}{\bibfnamefont{J.~H.} \bibnamefont{Cole}}, \bibnamefont{and}
  \bibinfo{author}{\bibfnamefont{L.~C.~L.} \bibnamefont{Hollenberg}},
  \bibinfo{journal}{Nature Physics} \textbf{\bibinfo{volume}{2}},
  \bibinfo{pages}{856} (\bibinfo{year}{2006}).

\bibitem[{\citenamefont{Hartmann et~al.}(2006)\citenamefont{Hartmann, Brandao,
  and Plenio}}]{Hartmann:2006}
\bibinfo{author}{\bibfnamefont{M.~J.} \bibnamefont{Hartmann}},
  \bibinfo{author}{\bibfnamefont{F.~G. S.~L.} \bibnamefont{Brandao}},
  \bibnamefont{and} \bibinfo{author}{\bibfnamefont{M.~B.}
  \bibnamefont{Plenio}}, \bibinfo{journal}{Nature Physics}
  \textbf{\bibinfo{volume}{2}}, \bibinfo{pages}{849} (\bibinfo{year}{2006}).

\bibitem[{\citenamefont{Chang et~al.}(2008)\citenamefont{Chang, Gritsev,
  Morigi, Vuletic, Lukin, and Demler}}]{Chang:2008}
\bibinfo{author}{\bibfnamefont{D.~E.} \bibnamefont{Chang}},
  \bibinfo{author}{\bibfnamefont{V.}~\bibnamefont{Gritsev}},
  \bibinfo{author}{\bibfnamefont{G.}~\bibnamefont{Morigi}},
  \bibinfo{author}{\bibfnamefont{V.}~\bibnamefont{Vuletic}},
  \bibinfo{author}{\bibfnamefont{M.~D.} \bibnamefont{Lukin}}, \bibnamefont{and}
  \bibinfo{author}{\bibfnamefont{E.~A.} \bibnamefont{Demler}},
  \bibinfo{journal}{Nature Physics} \textbf{\bibinfo{volume}{4}},
  \bibinfo{pages}{884} (\bibinfo{year}{2008}).

\bibitem[{\citenamefont{Schmidt and Blatter}(2010)}]{Schmidt:2010}
\bibinfo{author}{\bibfnamefont{S.}~\bibnamefont{Schmidt}} \bibnamefont{and}
  \bibinfo{author}{\bibfnamefont{G.}~\bibnamefont{Blatter}},
  \bibinfo{journal}{Physical Review Letters} \textbf{\bibinfo{volume}{104}},
  \bibinfo{pages}{216402} (\bibinfo{year}{2010}).
  
  
  
\bibitem[{\citenamefont{Carmichael et~al.}(1991)}]{cooper}
\bibinfo{author}{\bibfnamefont{N.}~\bibnamefont{Cooper}},
  \bibinfo{journal}{Advances in Physics} \textbf{\bibinfo{volume}{57}},
  \bibinfo{pages}{539} (\bibinfo{year}{2008}).
  
  \bibitem[{\citenamefont{Carmichael et~al.}(1991)}]{dalibard}
\bibinfo{author}{\bibfnamefont{J.}~\bibnamefont{Dalibard}}, 
  \bibinfo{author}{\bibfnamefont{et.~al.}},
  \bibinfo{journal}{Rev. Mod. Phys.} \textbf{\bibinfo{volume}{83}},
  \bibinfo{pages}{1523} (\bibinfo{year}{2011}).
  
    \bibitem[{\citenamefont{Carmichael et~al.}(1991)}]{bloch}
\bibinfo{author}{\bibfnamefont{I.}~\bibnamefont{Bloch}},
  \bibinfo{author}{\bibfnamefont{et.~al.}},
  \bibinfo{journal}{Rev. Mod. Phys.} \textbf{\bibinfo{volume}{80}},
  \bibinfo{pages}{885} (\bibinfo{year}{2008}).
  

\bibitem[{\citenamefont{Carusotto et~al.}(2009)\citenamefont{Carusotto, Gerace,
  Tureci, De~Liberato, Ciuti, and Imamo{\v g}lu}}]{Carusotto:2009}
\bibinfo{author}{\bibfnamefont{I.}~\bibnamefont{Carusotto}},
  \bibinfo{author}{\bibfnamefont{D.}~\bibnamefont{Gerace}},
  \bibinfo{author}{\bibfnamefont{H.}~\bibnamefont{Tureci}},
  \bibinfo{author}{\bibfnamefont{S.}~\bibnamefont{De~Liberato}},
  \bibinfo{author}{\bibfnamefont{C.}~\bibnamefont{Ciuti}}, \bibnamefont{and}
  \bibinfo{author}{\bibfnamefont{A.}~\bibnamefont{Imamo{\v g}lu}},
  \bibinfo{journal}{Physical Review Letters} \textbf{\bibinfo{volume}{103}},
  \bibinfo{pages}{033601} (\bibinfo{year}{2009}).

\bibitem[{\citenamefont{Tomadin et~al.}(2010)\citenamefont{Tomadin,
  Giovannetti, Fazio, Gerace, Carusotto, Tureci, and
  Imamoglu}}]{Tomadin:2010ul}
\bibinfo{author}{\bibfnamefont{A.}~\bibnamefont{Tomadin}},
  \bibinfo{author}{\bibfnamefont{V.}~\bibnamefont{Giovannetti}},
  \bibinfo{author}{\bibfnamefont{R.}~\bibnamefont{Fazio}},
  \bibinfo{author}{\bibfnamefont{D.}~\bibnamefont{Gerace}},
  \bibinfo{author}{\bibfnamefont{I.}~\bibnamefont{Carusotto}},
  \bibinfo{author}{\bibfnamefont{H.}~\bibnamefont{Tureci}}, \bibnamefont{and}
  \bibinfo{author}{\bibfnamefont{A.}~\bibnamefont{Imamoglu}},
  \bibinfo{journal}{Physical Review A} \textbf{\bibinfo{volume}{81}},
  \bibinfo{pages}{061801} (\bibinfo{year}{2010}).

\bibitem[{\citenamefont{Hafezi et~al.}(2011{\natexlab{a}})\citenamefont{Hafezi,
  Chang, Gritsev, Demler, and Lukin}}]{Hafezi:2011p36115}
\bibinfo{author}{\bibfnamefont{M.}~\bibnamefont{Hafezi}},
  \bibinfo{author}{\bibfnamefont{D.}~\bibnamefont{Chang}},
  \bibinfo{author}{\bibfnamefont{V.}~\bibnamefont{Gritsev}},
  \bibinfo{author}{\bibfnamefont{E.}~\bibnamefont{Demler}}, \bibnamefont{and}
  \bibinfo{author}{\bibfnamefont{M.}~\bibnamefont{Lukin}},
  \bibinfo{journal}{EPL (Europhysics Letters)} \textbf{\bibinfo{volume}{94}},
  \bibinfo{pages}{54006} (\bibinfo{year}{2011}{\natexlab{a}}).

\bibitem[{\citenamefont{Nunnenkamp et~al.}(2011)\citenamefont{Nunnenkamp, Koch,
  and Girvin}}]{Nunnenkamp:2011dm}
\bibinfo{author}{\bibfnamefont{A.}~\bibnamefont{Nunnenkamp}},
  \bibinfo{author}{\bibfnamefont{J.}~\bibnamefont{Koch}}, \bibnamefont{and}
  \bibinfo{author}{\bibfnamefont{S.~M.} \bibnamefont{Girvin}},
  \bibinfo{journal}{New Journal of Physics} \textbf{\bibinfo{volume}{13}},
  \bibinfo{pages}{095008} (\bibinfo{year}{2011}).

\bibitem[{\citenamefont{Nissen et~al.}(2012)\citenamefont{Nissen, Schmidt,
  Biondi, Blatter, Tureci, and Keeling}}]{Nissen:2012}
\bibinfo{author}{\bibfnamefont{F.}~\bibnamefont{Nissen}},
  \bibinfo{author}{\bibfnamefont{S.}~\bibnamefont{Schmidt}},
  \bibinfo{author}{\bibfnamefont{M.}~\bibnamefont{Biondi}},
  \bibinfo{author}{\bibfnamefont{G.}~\bibnamefont{Blatter}},
  \bibinfo{author}{\bibfnamefont{H.~E.} \bibnamefont{Tureci}},
  \bibnamefont{and} \bibinfo{author}{\bibfnamefont{J.}~\bibnamefont{Keeling}},
  \bibinfo{journal}{Physical Review Letters} \textbf{\bibinfo{volume}{108}},
  \bibinfo{pages}{233603} (\bibinfo{year}{2012}).

\bibitem[{\citenamefont{Schir{\'o} et~al.}(2012)\citenamefont{Schir{\'o},
  Bordyuh, {\"O}ztop, and Tureci}}]{schiro:2012}
\bibinfo{author}{\bibfnamefont{M.}~\bibnamefont{Schir{\'o}}},
  \bibinfo{author}{\bibfnamefont{M.}~\bibnamefont{Bordyuh}},
  \bibinfo{author}{\bibfnamefont{B.}~\bibnamefont{{\"O}ztop}},
  \bibnamefont{and} \bibinfo{author}{\bibfnamefont{H.}~\bibnamefont{Tureci}},
  \bibinfo{journal}{Physical Review Letters} \textbf{\bibinfo{volume}{109}},
  \bibinfo{pages}{053601} (\bibinfo{year}{2012}).

\bibitem[{\citenamefont{Koch et~al.}(2010)\citenamefont{Koch, Houck, Hur, and
  Girvin}}]{Koch:2010}
\bibinfo{author}{\bibfnamefont{J.}~\bibnamefont{Koch}},
  \bibinfo{author}{\bibfnamefont{A.~A.} \bibnamefont{Houck}},
  \bibinfo{author}{\bibfnamefont{K.~L.} \bibnamefont{Hur}}, \bibnamefont{and}
  \bibinfo{author}{\bibfnamefont{S.~M.} \bibnamefont{Girvin}},
  \bibinfo{journal}{Physical Review A} \textbf{\bibinfo{volume}{82}},
  \bibinfo{pages}{043811} (\bibinfo{year}{2010}).

\bibitem[{\citenamefont{Hafezi et~al.}(2011{\natexlab{b}})\citenamefont{Hafezi,
  Demler, Lukin, and Taylor}}]{Hafezi:2011delay}
\bibinfo{author}{\bibfnamefont{M.}~\bibnamefont{Hafezi}},
  \bibinfo{author}{\bibfnamefont{E.~A.} \bibnamefont{Demler}},
  \bibinfo{author}{\bibfnamefont{M.~D.} \bibnamefont{Lukin}}, \bibnamefont{and}
  \bibinfo{author}{\bibfnamefont{J.~M.} \bibnamefont{Taylor}},
  \bibinfo{journal}{Nature Physics} \textbf{\bibinfo{volume}{7}},
  \bibinfo{pages}{907} (\bibinfo{year}{2011}{\natexlab{b}}).

\bibitem[{\citenamefont{Hafezi and Rabl}(2012)}]{Hafezi:2011ui}
\bibinfo{author}{\bibfnamefont{M.}~\bibnamefont{Hafezi}} \bibnamefont{and}
  \bibinfo{author}{\bibfnamefont{P.}~\bibnamefont{Rabl}},
  \bibinfo{journal}{Optics Express} \textbf{\bibinfo{volume}{20}},
  \bibinfo{pages}{7672} (\bibinfo{year}{2012}).

\bibitem[{\citenamefont{Umucalilar and Carusotto}(2011)}]{Umucalilar:2011}
\bibinfo{author}{\bibfnamefont{R.~O.} \bibnamefont{Umucalilar}}
  \bibnamefont{and}
  \bibinfo{author}{\bibfnamefont{I.}~\bibnamefont{Carusotto}},
  \bibinfo{journal}{Physical Review A} \textbf{\bibinfo{volume}{84}},
  \bibinfo{pages}{043804} (\bibinfo{year}{2011}).

\bibitem[{\citenamefont{Cho et~al.}(2008)\citenamefont{Cho, Angelakis, and
  Bose}}]{Cho:2008}
\bibinfo{author}{\bibfnamefont{J.}~\bibnamefont{Cho}},
  \bibinfo{author}{\bibfnamefont{D.}~\bibnamefont{Angelakis}},
  \bibnamefont{and} \bibinfo{author}{\bibfnamefont{S.}~\bibnamefont{Bose}},
  \bibinfo{journal}{Physical Review Letters} \textbf{\bibinfo{volume}{101}},
  \bibinfo{pages}{246809} (\bibinfo{year}{2008}).

\bibitem[{\citenamefont{Umucalilar and Carusotto}(2012)}]{Umucalilar:2011b}
\bibinfo{author}{\bibfnamefont{R.~O.} \bibnamefont{Umucalilar}}
  \bibnamefont{and}
  \bibinfo{author}{\bibfnamefont{I.}~\bibnamefont{Carusotto}},
  \bibinfo{journal}{Phys Rev Lett} \textbf{\bibinfo{volume}{108}},
  \bibinfo{pages}{206809} (\bibinfo{year}{2012}).

\bibitem[{\citenamefont{Hayward and Martin}(2012)}]{Hayward:2012}
\bibinfo{author}{\bibfnamefont{A.}~\bibnamefont{Hayward}} \bibnamefont{and}
  \bibinfo{author}{\bibfnamefont{A.~M.} \bibnamefont{Martin}},
  \bibinfo{journal}{Physical Review Letters} \textbf{\bibinfo{volume}{108}},
  \bibinfo{pages}{223602} (\bibinfo{year}{2012}).

\bibitem[{\citenamefont{Birnbaum et~al.}(2005)\citenamefont{Birnbaum, Boca,
  Miller, Boozer, Northup, and Kimble}}]{Kimble:2005}
\bibinfo{author}{\bibfnamefont{K.~M.} \bibnamefont{Birnbaum}},
  \bibinfo{author}{\bibfnamefont{A.}~\bibnamefont{Boca}},
  \bibinfo{author}{\bibfnamefont{R.}~\bibnamefont{Miller}},
  \bibinfo{author}{\bibfnamefont{A.~D.} \bibnamefont{Boozer}},
  \bibinfo{author}{\bibfnamefont{T.~E.} \bibnamefont{Northup}},
  \bibnamefont{and} \bibinfo{author}{\bibfnamefont{H.~J.}
  \bibnamefont{Kimble}}, \bibinfo{journal}{Nature}
  \textbf{\bibinfo{volume}{436}}, \bibinfo{pages}{87} (\bibinfo{year}{2005}).

\bibitem[{\citenamefont{Englund et~al.}(2007)\citenamefont{Englund, Faraon,
  Fushman, Stoltz, Petroff, and Vuckovic}}]{Englund:2007}
\bibinfo{author}{\bibfnamefont{D.}~\bibnamefont{Englund}},
  \bibinfo{author}{\bibfnamefont{A.}~\bibnamefont{Faraon}},
  \bibinfo{author}{\bibfnamefont{I.}~\bibnamefont{Fushman}},
  \bibinfo{author}{\bibfnamefont{N.}~\bibnamefont{Stoltz}},
  \bibinfo{author}{\bibfnamefont{P.}~\bibnamefont{Petroff}}, \bibnamefont{and}
  \bibinfo{author}{\bibfnamefont{J.}~\bibnamefont{Vuckovic}},
  \bibinfo{journal}{Nature} \textbf{\bibinfo{volume}{450}},
  \bibinfo{pages}{857} (\bibinfo{year}{2007}).

\bibitem[{\citenamefont{Srinivasan and Painter}(2007)}]{Srinivasan:2007p4061}
\bibinfo{author}{\bibfnamefont{K.}~\bibnamefont{Srinivasan}} \bibnamefont{and}
  \bibinfo{author}{\bibfnamefont{O.}~\bibnamefont{Painter}},
  \bibinfo{journal}{Nature} \textbf{\bibinfo{volume}{450}},
  \bibinfo{pages}{862} (\bibinfo{year}{2007}).

\bibitem[{\citenamefont{Schoelkopf and Girvin}(2008)}]{Schoelkopf:2008p8712}
\bibinfo{author}{\bibfnamefont{R.~J.} \bibnamefont{Schoelkopf}}
  \bibnamefont{and} \bibinfo{author}{\bibfnamefont{S.~M.}
  \bibnamefont{Girvin}}, \bibinfo{journal}{Nature}
  \textbf{\bibinfo{volume}{451}}, \bibinfo{pages}{664} (\bibinfo{year}{2008}).

\bibitem[{\citenamefont{You and Nori}(2011)}]{You:2011jj}
\bibinfo{author}{\bibfnamefont{J.~Q.} \bibnamefont{You}} \bibnamefont{and}
  \bibinfo{author}{\bibfnamefont{F.}~\bibnamefont{Nori}},
  \bibinfo{journal}{Nature} \textbf{\bibinfo{volume}{474}},
  \bibinfo{pages}{589} (\bibinfo{year}{2011}).

\bibitem[{\citenamefont{Carmichael et~al.}(1991)\citenamefont{Carmichael,
  Brecha, and Rice}}]{Carmichael:1991}
\bibinfo{author}{\bibfnamefont{H.}~\bibnamefont{Carmichael}},
  \bibinfo{author}{\bibfnamefont{R.}~\bibnamefont{Brecha}}, \bibnamefont{and}
  \bibinfo{author}{\bibfnamefont{P.}~\bibnamefont{Rice}},
  \bibinfo{journal}{Optics Communications} \textbf{\bibinfo{volume}{82}},
  \bibinfo{pages}{73} (\bibinfo{year}{1991}).

  \bibitem[{\citenamefont{Carmichael et~al.}(1991)}]{mekhov:phys}
\bibinfo{author}{\bibfnamefont{I.}~\bibnamefont{Mekhov}}, 
\bibinfo{author}{\bibfnamefont{C.}~\bibnamefont{Maschler}}, 
\bibnamefont{and}
  \bibinfo{author}{\bibfnamefont{H.}~\bibnamefont{Ritsch}},
  \bibinfo{journal}{ Nature Physics } \textbf{\bibinfo{volume}{3}},
  \bibinfo{pages}{319} (\bibinfo{year}{2007}).



  \bibitem[{\citenamefont{Carmichael et~al.}(1991)}]{chen}
\bibinfo{author}{\bibfnamefont{W.}~\bibnamefont{Chen}}, 
\bibinfo{author}{\bibfnamefont{D.}~\bibnamefont{Meiser}},\bibnamefont{and}
  \bibinfo{author}{\bibfnamefont{P.}~\bibnamefont{Meystre}},
  \bibinfo{journal}{ Phys. Rev. A } \textbf{\bibinfo{volume}{75}},
  \bibinfo{pages}{023812} (\bibinfo{year}{2007}).
  



\bibitem[{\citenamefont{Carmichael et~al.}(1991)}]{mekhov}
\bibinfo{author}{\bibfnamefont{I.}~\bibnamefont{Mekhov}}, \bibnamefont{and}
  \bibinfo{author}{\bibfnamefont{H.}~\bibnamefont{Ritsch}},
  \bibinfo{journal}{ J. Phys. B: At. Mol. Opt. Phys. } \textbf{\bibinfo{volume}{45}},
  \bibinfo{pages}{102001} (\bibinfo{year}{2012}).
  
  

\bibitem[{\citenamefont{S{\o}rensen et~al.}(2005)\citenamefont{S{\o}rensen,
  Demler, and Lukin}}]{sorensen}
\bibinfo{author}{\bibfnamefont{A.~S.} \bibnamefont{S{\o}rensen}},
  \bibinfo{author}{\bibfnamefont{E.}~\bibnamefont{Demler}}, \bibnamefont{and}
  \bibinfo{author}{\bibfnamefont{M.~D.} \bibnamefont{Lukin}},
  \bibinfo{journal}{Physical Review Letters} \textbf{\bibinfo{volume}{94}},
  \bibinfo{pages}{086803} (\bibinfo{year}{2005}).

\bibitem[{\citenamefont{Hafezi et~al.}(2007)\citenamefont{Hafezi, S{\o}rensen,
  Demler, and Lukin}}]{Hafezi:PRA2007}
\bibinfo{author}{\bibfnamefont{M.}~\bibnamefont{Hafezi}},
  \bibinfo{author}{\bibfnamefont{A.~S.} \bibnamefont{S{\o}rensen}},
  \bibinfo{author}{\bibfnamefont{E.}~\bibnamefont{Demler}}, \bibnamefont{and}
  \bibinfo{author}{\bibfnamefont{M.~D.} \bibnamefont{Lukin}},
  \bibinfo{journal}{Physical Review A} \textbf{\bibinfo{volume}{76}},
  \bibinfo{pages}{023613} (\bibinfo{year}{2007}).

\bibitem[{\citenamefont{Dalibard et~al.}(1992)\citenamefont{Dalibard, Castin,
  and M~olmer}}]{Dalibard:1992}
\bibinfo{author}{\bibfnamefont{J.}~\bibnamefont{Dalibard}},
  \bibinfo{author}{\bibfnamefont{Y.}~\bibnamefont{Castin}}, \bibnamefont{and}
  \bibinfo{author}{\bibfnamefont{K.}~\bibnamefont{Molmer}},
  \bibinfo{journal}{Physical Review Letters} \textbf{\bibinfo{volume}{68}},
  \bibinfo{pages}{580} (\bibinfo{year}{1992}).

\bibitem[{\citenamefont{Carmichael}(2007)}]{carmichael:book}
\bibinfo{author}{\bibfnamefont{H.}~\bibnamefont{Carmichael}},
  \emph{\bibinfo{title}{{Statistical Methods in Quantum Optics: Non-classical
  fields}}} (\bibinfo{publisher}{Springer}, \bibinfo{year}{2007}).

\bibitem[{\citenamefont{K~artner and Haus}(1993)}]{kartner93}
\bibinfo{author}{\bibfnamefont{F.~X.} \bibnamefont{K~artner}} \bibnamefont{and}
  \bibinfo{author}{\bibfnamefont{H.~A.} \bibnamefont{Haus}},
  \bibinfo{journal}{Physical Review A} \textbf{\bibinfo{volume}{48}},
  \bibinfo{pages}{2361} (\bibinfo{year}{1993}).

\bibitem[{\citenamefont{Bajcsy et~al.}(2009)\citenamefont{Bajcsy, Hofferberth,
  Balic, Peyronel, Hafezi, Zibrov, Vuletic, and Lukin}}]{Bajcsy:2009p6498}
\bibinfo{author}{\bibfnamefont{M.}~\bibnamefont{Bajcsy}},
  \bibinfo{author}{\bibfnamefont{S.}~\bibnamefont{Hofferberth}},
  \bibinfo{author}{\bibfnamefont{V.}~\bibnamefont{Balic}},
  \bibinfo{author}{\bibfnamefont{T.}~\bibnamefont{Peyronel}},
  \bibinfo{author}{\bibfnamefont{M.}~\bibnamefont{Hafezi}},
  \bibinfo{author}{\bibfnamefont{A.}~\bibnamefont{Zibrov}},
  \bibinfo{author}{\bibfnamefont{V.}~\bibnamefont{Vuletic}}, \bibnamefont{and}
  \bibinfo{author}{\bibfnamefont{M.}~\bibnamefont{Lukin}},
  \bibinfo{journal}{Physical Review Letters} \textbf{\bibinfo{volume}{102}},
  \bibinfo{pages}{203902} (\bibinfo{year}{2009}).

\bibitem[{\citenamefont{Fushman et~al.}(2008)\citenamefont{Fushman, Englund,
  Faraon, Stoltz, Petroff, and Vuckovic}}]{Fushman:2008}
\bibinfo{author}{\bibfnamefont{I.}~\bibnamefont{Fushman}},
  \bibinfo{author}{\bibfnamefont{D.}~\bibnamefont{Englund}},
  \bibinfo{author}{\bibfnamefont{A.}~\bibnamefont{Faraon}},
  \bibinfo{author}{\bibfnamefont{N.}~\bibnamefont{Stoltz}},
  \bibinfo{author}{\bibfnamefont{P.}~\bibnamefont{Petroff}}, \bibnamefont{and}
  \bibinfo{author}{\bibfnamefont{J.}~\bibnamefont{Vuckovic}},
  \bibinfo{journal}{Science} \textbf{\bibinfo{volume}{320}},
  \bibinfo{pages}{769} (\bibinfo{year}{2008}).

\bibitem[{\citenamefont{Ga{\"e}tan et~al.}(2009)\citenamefont{Ga{\"e}tan,
  Miroshnychenko, Wilk, Chotia, Viteau, Comparat, Pillet, Browaeys, and
  Grangier}}]{Gaetan:2009}
\bibinfo{author}{\bibfnamefont{A.}~\bibnamefont{Ga{\"e}tan}},
  \bibinfo{author}{\bibfnamefont{Y.}~\bibnamefont{Miroshnychenko}},
  \bibinfo{author}{\bibfnamefont{T.}~\bibnamefont{Wilk}},
  \bibinfo{author}{\bibfnamefont{A.}~\bibnamefont{Chotia}},
  \bibinfo{author}{\bibfnamefont{M.}~\bibnamefont{Viteau}},
  \bibinfo{author}{\bibfnamefont{D.}~\bibnamefont{Comparat}},
  \bibinfo{author}{\bibfnamefont{P.}~\bibnamefont{Pillet}},
  \bibinfo{author}{\bibfnamefont{A.}~\bibnamefont{Browaeys}}, \bibnamefont{and}
  \bibinfo{author}{\bibfnamefont{P.}~\bibnamefont{Grangier}},
  \bibinfo{journal}{Nature Physics} \textbf{\bibinfo{volume}{5}},
  \bibinfo{pages}{115} (\bibinfo{year}{2009}).

\bibitem[{\citenamefont{Urban et~al.}(2009)\citenamefont{Urban, Johnson,
  Henage, Isenhower, Yavuz, Walker, and Saffman}}]{Urban:2009}
\bibinfo{author}{\bibfnamefont{E.}~\bibnamefont{Urban}},
  \bibinfo{author}{\bibfnamefont{T.~A.} \bibnamefont{Johnson}},
  \bibinfo{author}{\bibfnamefont{T.}~\bibnamefont{Henage}},
  \bibinfo{author}{\bibfnamefont{L.}~\bibnamefont{Isenhower}},
  \bibinfo{author}{\bibfnamefont{D.~D.} \bibnamefont{Yavuz}},
  \bibinfo{author}{\bibfnamefont{T.~G.} \bibnamefont{Walker}},
  \bibnamefont{and} \bibinfo{author}{\bibfnamefont{M.}~\bibnamefont{Saffman}},
  \bibinfo{journal}{Nature Physics} \textbf{\bibinfo{volume}{5}},
  \bibinfo{pages}{110} (\bibinfo{year}{2009}).

\bibitem[{\citenamefont{Peyronel et~al.}(2012)\citenamefont{Peyronel,
  Firstenberg, Liang, Hofferberth, Gorshkov, Pohl, Lukin, and
  Vuletic}}]{Peyronel:2012}
\bibinfo{author}{\bibfnamefont{T.}~\bibnamefont{Peyronel}},
  \bibinfo{author}{\bibfnamefont{O.}~\bibnamefont{Firstenberg}},
  \bibinfo{author}{\bibfnamefont{Q.~Y.} \bibnamefont{Liang}},
  \bibinfo{author}{\bibfnamefont{S.}~\bibnamefont{Hofferberth}},
  \bibinfo{author}{\bibfnamefont{A.~V.} \bibnamefont{Gorshkov}},
  \bibinfo{author}{\bibfnamefont{T.}~\bibnamefont{Pohl}},
  \bibinfo{author}{\bibfnamefont{M.~D.} \bibnamefont{Lukin}}, \bibnamefont{and}
  \bibinfo{author}{\bibfnamefont{V.}~\bibnamefont{Vuletic}},
  \bibinfo{journal}{Nature} \textbf{\bibinfo{volume}{488}}, \bibinfo{pages}{57}
  (\bibinfo{year}{2012}).

\bibitem[{\citenamefont{Lang et~al.}(2011)\citenamefont{Lang, Bozyigit,
  Eichler, Steffen, Fink, Abdumalikov, Baur, Filipp, da~Silva, Blais
  et~al.}}]{Lang:2011}
\bibinfo{author}{\bibfnamefont{C.}~\bibnamefont{Lang}},
  \bibinfo{author}{\bibfnamefont{D.}~\bibnamefont{Bozyigit}},
  \bibinfo{author}{\bibfnamefont{C.}~\bibnamefont{Eichler}},
  \bibinfo{author}{\bibfnamefont{L.}~\bibnamefont{Steffen}},
  \bibinfo{author}{\bibfnamefont{J.}~\bibnamefont{Fink}},
  \bibinfo{author}{\bibfnamefont{A.}~\bibnamefont{Abdumalikov}},
  \bibinfo{author}{\bibfnamefont{M.}~\bibnamefont{Baur}},
  \bibinfo{author}{\bibfnamefont{S.}~\bibnamefont{Filipp}},
  \bibinfo{author}{\bibfnamefont{M.}~\bibnamefont{da~Silva}},
  \bibinfo{author}{\bibfnamefont{A.}~\bibnamefont{Blais}},
  \bibnamefont{et~al.}, \bibinfo{journal}{Physical Review Letters}
  \textbf{\bibinfo{volume}{106}}, \bibinfo{pages}{243601}
  (\bibinfo{year}{2011}).

\bibitem[{\citenamefont{Pohl et~al.}(2010)\citenamefont{Pohl, Demler, and
  Lukin}}]{Pohl:2010}
\bibinfo{author}{\bibfnamefont{T.}~\bibnamefont{Pohl}},
  \bibinfo{author}{\bibfnamefont{E.}~\bibnamefont{Demler}}, \bibnamefont{and}
  \bibinfo{author}{\bibfnamefont{M.~D.} \bibnamefont{Lukin}},
  \bibinfo{journal}{Physical Review Letters} \textbf{\bibinfo{volume}{104}},
  \bibinfo{pages}{043002} (\bibinfo{year}{2010}).

\bibitem[{\citenamefont{Fleischhauer and Lukin}(2000)}]{Fleischhauer:2000kx}
\bibinfo{author}{\bibfnamefont{M.}~\bibnamefont{Fleischhauer}}
  \bibnamefont{and} \bibinfo{author}{\bibfnamefont{M.~D.} \bibnamefont{Lukin}},
  \bibinfo{journal}{Physical Review Letters} \textbf{\bibinfo{volume}{84}},
  \bibinfo{pages}{5094} (\bibinfo{year}{2000}).

\bibitem[{\citenamefont{Andre et~al.}(2005)\citenamefont{Andre, Bajcsy, Zibrov,
  and Lukin}}]{andre:NLO}
\bibinfo{author}{\bibfnamefont{A.}~\bibnamefont{Andre}},
  \bibinfo{author}{\bibfnamefont{M.}~\bibnamefont{Bajcsy}},
  \bibinfo{author}{\bibfnamefont{A.~S.} \bibnamefont{Zibrov}},
  \bibnamefont{and} \bibinfo{author}{\bibfnamefont{M.~D.} \bibnamefont{Lukin}},
  \bibinfo{journal}{Physical Review Letters} \textbf{\bibinfo{volume}{94}},
  \bibinfo{pages}{063902} (\bibinfo{year}{2005}).

\bibitem[{\citenamefont{Diehl et~al.}(2008)\citenamefont{Diehl, Micheli,
  Kantian, Kraus, B{\"u}chler, and Zoller}}]{Diehl:2008}
\bibinfo{author}{\bibfnamefont{S.}~\bibnamefont{Diehl}},
  \bibinfo{author}{\bibfnamefont{A.}~\bibnamefont{Micheli}},
  \bibinfo{author}{\bibfnamefont{A.}~\bibnamefont{Kantian}},
  \bibinfo{author}{\bibfnamefont{B.}~\bibnamefont{Kraus}},
  \bibinfo{author}{\bibfnamefont{H.~P.} \bibnamefont{B{\"u}chler}},
  \bibnamefont{and} \bibinfo{author}{\bibfnamefont{P.}~\bibnamefont{Zoller}},
  \bibinfo{journal}{Nature Physics} \textbf{\bibinfo{volume}{4}},
  \bibinfo{pages}{878} (\bibinfo{year}{2008}).

\end{thebibliography}
\end{document}